\documentclass{aa}  

\usepackage{graphicx}
\usepackage{txfonts}
\usepackage{hyperref}
\usepackage{siunitx}
\usepackage{xcolor}
\usepackage{amsmath}
\usepackage{longtable}
\usepackage{placeins}

\definecolor{frcolor}{rgb}{0,0.5,0}

\newcommand{\msun}{M_\odot}
\newcommand{\rsun}{R_\odot}

\begin{document}

   \title{The asymmetry of white dwarf double detonations and the observed scatter around the Phillips relation}
    \titlerunning{The asymmetry of white dwarf double detonations}
    \authorrunning{Holas et al.}

   \author{
        Alexander Holas\inst{1}
        \and
        Friedrich K. Röpke\inst{1,2}
        \and
        Rüdiger Pakmor\inst{3}
        \and
        Fionntan P. Callan\inst{4}
        \and
        Josh Pollin\inst{4}
        \and
        Stuart A. Sim\inst{4}
        \and
        Christine E. Collins\inst{5}
        \and
        Luke J. Shingles\inst{6}
        \and
        Javier Moran-Fraile\inst{1,2}
    }

   \institute{
        Heidelberger Institut für Theoretische Studien,
        Schloss-Wolfsbrunnenweg 35, 69118 Heidelberg, Germany\\
        \email{alexander.holas@mailbox.org}
        \and
        Zentrum f\"ur Astronomie der Universit\"at Heidelberg,
        Institut f\"ur Theoretische Astrophysik, 
        Philosophenweg 12,
        69120 Heidelberg, Germany
        \and
        Max-Planck-Institut f\"ur Astrophysik,
        Karl-Schwarzschild-Str. 1, 85748 Garching,
        Germany
        \and
        School of Mathematics and Physics, Queen's University Belfast, University   Road, Belfast BT7 1NN, UK 
        \and
        School of Physics, Trinity College Dublin, The University of Dublin, Dublin 2, Ireland
        \and
        GSI Helmholtzzentrum f\"{u}r Schwerionenforschung, Planckstraße 1, 64291 Darmstadt, Germany
    }

   \date{Received September 15, 1996; accepted March 16, 1997}
   
\abstract{
Recent Type Ia supernova (SN Ia) simulations featuring a double detonation scenario have managed to reproduce the overall trend of the Phillips relation reasonably well. However, most, if not all, multidimensional simulations struggle to reproduce the scatter of observed SNe around this relation, exceeding it substantially.
}{
In this study, we investigate whether the excessive scatter around the Phillips relation can be caused by an off-center ignition of the carbon-oxygen (CO) core in the double detonation scenario and if this can help constrain possible SN Ia explosion channels.
}{
We simulated the detonation of three different initial CO white dwarfs of $0.9$, $1.0$, and $1.1\,\msun$, artificially ignited at systematically offset locations using the \textsc{Arepo} code. After nucleosynthetic postprocessing, we generated synthetic observables using the \textsc{Artis} code and compared these results against observational data and models of other works.
}{
We find that our simulations produce synthetic observables well within the range of the observed data in terms of viewing angle scatter. The majority of the viewing angle variability seems to be caused by line blanketing in the blue wavelengths of intermediate-mass elements and lighter iron-group elements, which are asymmetrically distributed in the outer layers of the ashes.
}{
Our results suggest that although the off-center ignition of the CO introduces substantial line of sight effects, it is not responsible for the excessive viewing angle scatter observed in other models. Instead, this effect seems to be caused by the detonation ashes from the rather massive helium (He) shells in current state-of-the-art models. Further reducing the He-shell masses of double detonation progenitors may be able to alleviate this issue and yield observables that reproduce the Phillips relation.
}


   \maketitle

\section{Introduction}\label{sec:intro}

Type Ia supernovae (SNe Ia) explosions became widely known for their use as cosmological distance indicators, aiding in the discovery of the accelerated expansion of the Universe \citep{riess1998a,schmidt1998a,perlmutter1999a}.
Although many subtypes of SNe Ia have been discovered and classified in recent years (see, e.g., \citealt{taubenberger2017a} for an overview), only one type of SN Ia has been widely used as a distance indicator in cosmology, generally referred to as ``normal'' Type Ia.
What sets this apart from other SN Ia subtypes is that its peak B band magnitude, $M_\mathrm{B,max}$, was found to be correlated to its decline rate, $\Delta \mathrm{m}_{15}(\mathrm{B})$, in the B band \citep{phillips1993a}, the so called Phillips relation.\footnote{While we focus on the B band here, similar relations also exists for the peak magnitude of other bands; \citet{phillips1993a} originally found a relation for the B, V, and I band.}
This empirical relation allowed for the standardization of SNe Ia and their use as distance indicators, yet it still lacks a physical explanation.
Despite their importance for fields such as cosmology, the precise nature of the SN Ia explosion mechanism and their progenitor systems remains equivocal.
It is commonly agreed that most if not all normal SNe Ia are caused by the thermonuclear explosion of a degenerate carbon-oxygen (CO) white dwarf (WD, \citealt{hoyle1960a}) that in some way interacts with a binary companion.
This interaction can occur in a single-degenerate \citep{whelan1973a} or a double-degenerate \citep{iben1984a} scenario, with an ongoing debate as to which of these scenarios is the dominant scenario for normal SNe Ia. 
Likewise, it is not yet clear whether normal SNe Ia are caused by explosions of near-Chandrasekhar mass ($M_\mathrm{Ch}$) or sub-$M_\mathrm{Ch}$ WDs (see \citealt{liu2023a} for an extensive overview).
In this work, we focus on sub-$M_\mathrm{Ch}$ progenitors as the source of normal SN Ia, as models of such scenarios have been shown to reproduce observations effectively \citep{sim2010a,blondin2017a,shen2018a,shen2021a} and have the ability to reproduce the Phillips relation rather well \citep{shen2021a}, at least its mean trend, by varying the mass of the exploding WD.
Although sub-$M_\mathrm{Ch}$ WDs can explode through various mechanisms, the double detonation process is of particular interest.
In short, this scenario involves a first detonation of a helium (He) layer on top of a CO WD.
The initial explosion generates a shockwave that penetrates into the CO core, causing the CO fuel to compress and ignite at the convergence point.
Following this, a subsequent detonation combusts the CO core, releasing the majority of the explosion's energy and creating most of the heavy elements found in SNe Ia \citep{nomoto1980a,taam1980a,livne1990a,hoeflich1996a,nugent1997a}.
This double detonation mechanism is called the converging shock mechanism; there exist other mechanisms such as edge-lit or scissor mechanisms (e.g., \citealt{gronow2021a}), but for the scope of this work, we focus on the converging shock scenario.
In this process, the He detonation predominantly produces intermediate-mass elements (IMEs), whereas the CO detonation produces the bulk of the iron-group elements (IGEs).

Although recent double detonation simulations are quite successful in reproducing observed SN Ia quantities, many state-of-the-art explosion models \citep{gronow2021a,collins2022a,shen2021b,pollin2024a} struggle in matching the observed scatter around the Phillips relation\footnote{Throughout this work, ``scatter around the Phillips relation'' refers to the dispersion around the mean $M_\mathrm{B,max}-\Delta\mathrm{m}_{15}(\mathrm{B})$ trend for a given model or observational dataset; the mean trend of the model itself can vary from the observed mean trend due to modeling choices. We use phrases such as ``scatter,'' ``dispersion'' or ``variability'' interchangeably in this context.}, which is likely caused by a viewing angle dependence of the observables \citep{sim2013a,shen2021b,gronow2021a,collins2022a}; although there may be secondary parameters other than line of sight effects contributing to this scatter, we focus on viewing angle contributions as the driving factor in this study.
These results appear to be independent of the dimensionality of the radiative transfer (RT) calculations; both fully 3D calculations (e.g., \citealt{pollin2024a}) and 1D line of sight calculations (e.g., \citealt{boos2024b}) face challenges in this respect.
We note that recent work by \cite{boos2024b} has made strides in better matching observables in that respect, but discrepancies remain.
This conclusion is the main motivation for our study.
It is not clear whether this excessive scatter is caused solely by an off-center ignition of the detonation of the CO core in the converging shock.
Moreover, a generic asymmetry that is too strong could help exclude some models; the impact of the He-shell detonation strongly depends on the particular sub-$M_\mathrm{Ch}$ explosion model and consequently would make statements on the explosion asymmetry model-dependent.

To investigate how much an off-center CO detonation contributes to the viewing angle scatter of observed quantities, we conducted a systematic parameter study of idealized CO detonations ignited at various locations in the CO core. 
Here, in order to have full control over the ignition of the CO detonation, we neglected the He-shell and artificially ignited a CO detonation in the core at various radii.
We also closely compared our models to the self-consistent models of \citet{pakmor2022a,pollin2024a} that are more representative of an actual double detonation SN Ia.

In Section~\ref{sec:methods}, we describe the methods that we use in our simulations, i.e., the hydrodynamic explosion simulation, the nucleosynthetic post-processing, and the radiative transfer calculations.
In Section~\ref{sec:results} we present the results obtained for our simulations and compare them with observational data and other recent state-of-the-art simulations.
We discuss our main findings in Section~\ref{sec:discussion} and summarize our results in Section~\ref{sec:conclusion}.

\section{Methods}\label{sec:methods}
\subsection{Detonation simulation}
We simulated the explosion in 3D utilizing the \textsc{Arepo} code \citep{springel2010a,pakmor2011a,pakmor2016a,weinberger2020a} in a similar fashion to \cite{pakmor2022a}.
\textsc{Arepo} uses a second-order finite volume approach on an unstructured moving Voronoi mesh, solving for ideal magneto-hydrodynamics, self-gravity and an Helmholtz equation of state \citep{timmes2000a}.
The (de)refinement of cells is carried out in an explicit manner: cells that differ by more than a factor of two from a set target mass ($m_\mathrm{target} = 1\times10^{-6}\,\msun$) undergo (de)refinement; cells that are more than $10$ times larger than their smallest direct neighboring cell undergo refinement.
To self-consistently model the explosion, we include a reaction network using the JINA reaction rates \citep{cyburt2010a} and couple it to the hydrodynamics \citep{pakmor2012a,pakmor2022a,gronow2021a}.
The network covers 55 isotopes: n, p, $^4$He, $^{11}$B, $^{12-13}$C, $^{13-15}$N, $^{15-17}$O, $^{18}$F, $^{19-22}$Ne, $^{22-23}$Na, $^{23-26}$Mg, $^{25-27}$Al, $^{28-30}$Si, $^{29-31}$P, $^{31-33}$S, $^{33-35}$Cl, $^{36-39}$Ar, $^{39}$K, $^{40}$Ca, $^{43}$Sc, $^{44}$Ti, $^{47}$V, $^{48}$Cr, $^{51}$Mn, $^{52,56}$Fe, $^{55}$Co, and $^{56,58-59}$Ni. It is designed to properly model the energy release from He, C, O and Si burning.
Nuclear burning is activated only for cells that reach a sufficient temperature ($T > 2\times 10 ^7\,\mathrm{K}$) and density ($\rho > 1 \times 10^4\,\mathrm{g}\,\mathrm{cm}^{-3}$).
However, we also use a shock limiter for burning \citep[see][for a description]{pakmor2022a}.

\subsection{Simulation setup}
Our initial setup consisted of three individual WDs with equal-by-mass CO mixtures with a total mass of $0.9$, $1.0$, and $1.1\,\msun$. 
Here, the He-shell was deliberately excluded from consideration in order to avoid contamination of our study by its influence.
We denote these models as wd\_m090, wd\_m100, and wd\_m110, respectively.
The mass distribution was roughly inspired by the masses investigated by \citet{gronow2021a}.
These WDs were constructed by solving the equations for hydrostatic equilibrium using a uniform temperature of $5\times 10^5\,\mathrm{K}$ .
This yielded WDs of sizes $0.01$, $0.009$, and $0.007\,\rsun$, respectively, which were placed in a cubic box of size $10^{12}$\,cm with a uniform background density of $\rho = 10^{-5}$\,g\,cm$^{-3}$.
The resulting density profiles were then mapped to a 3D mesh using a HEALPix algorithm following \citet{pakmor2012a,ohlmann2017a}, and the internal energy of each cell was then calculated from the density and temperature using a Helmholtz equation of state.
After setting up the WDs in \textsc{Arepo}, we relaxed it, both to remove noise introduced by the mapping process and to ensure hydrostatic equilibrium with the correct density profile (see \citealt{pakmor2012a,ohlmann2017a,pakmor2022a}).
The relaxation duration was set to about ten dynamical timescales for each respective initial WD.
Furthermore, we checked the stability by letting the WDs evolve freely for another ${\sim}\,100\,\mathrm{s}$ in a separate simulation.
In all cases, the density profiles remain virtually unchanged, confirming the stability of our initial conditions.

These relaxed WDs were then artificially ignited by placing an ignition spot at a predefined location.
Each ignition spot contains $10^{-4}\,\msun$ at a temperature of $3\times10^9\,\mathrm{K}$, keeping the injected energy roughly constant for all ignition spots.
The locations at which we ignite the WDs roughly follow a $100\times2^n\,\mathrm{km}$ scheme, starting at the center and moving outward up to a density of $2\times10^6\,\mathrm{g}\,\mathrm{cm}^{-3}$, below which ignition of a detonation is considered unrealistic \citep{seitenzahl2009b}.
We designate each ignition spot as r<distance from center in km>, i.e., a $0.9\,\msun$ WD ignited $100\,\mathrm{km}$ off-center is referred to as wd\_m090\_r100.
An overview of the ignition spots and the respective densities at their position alongside their approximate radii can be found in Table \ref{tab:ignition}.

\begin{table}
    \centering 
    \caption{List of the densities at the center of each ignition spot and the radii of the ignition spots.}\label{tab:ignition}
    \begingroup
	\setlength{\tabcolsep}{6pt}
	\renewcommand{\arraystretch}{1.25} 
    \begin{tabular}{c | c c c}
    \hline\hline
     & \multicolumn{3}{c}{$\rho_\mathrm{ign}$ ($10^7$\,g\,cm$^{-3}$)}\\
    Ignition setup & wd\_m090 & wd\_m100 & wd\_m110 \\
    \hline
    r0              & 1.95 & 3.51 & 6.88 \\
    r100            & 1.94 & 3.51 & 6.85 \\
    r200            & 1.93 & 3.47 & 6.77 \\
    r400            & 1.89 & 3.38 & 6.52 \\
    r800            & 1.74 & 3.00 & 5.41 \\
    r1600           & 1.26 & 1.96 & 2.93 \\
    r2800 (r2400\tablefootmark{a})  & 0.79 & 0.66 & 0.65 \\
    \hline 
     & \multicolumn{3}{c}{$r_\mathrm{ign}$ (km)\tablefootmark{b}}\\
    \hline
    r0 & 135.0 & 110.7 & 88.7 \\
    r100 & 135.4 & 110.9 & 88.9 \\
    r200 & 135.4 & 111.3 & 88.8 \\
    r400 & 136.7 & 112.3 & 90.4 \\
    r800 & 139.6 & 116.5 & 95.8 \\
    r1600 & 155.3 & 135.4 & 118.5 \\
    r2800 (r2400\tablefootmark{a}) & 202.9 & 193.2 & 194.5 \\
    \hline
    \end{tabular}
    \endgroup
    \tablefoot{
        \tablefoottext{a}{Ignition spot location for wd\_m090.}
        \tablefoottext{b}{Approximate radius of the ignition spot calculated from the ignition spot volume.}
    }
\end{table}

For the outermost ignition spots, we deviated from the regular $2^n$ pattern (that is, the most off-center location is at 2400\,km and 2800\,km), as we found that placing our $10^{-4}\,\msun$ ``hot bubble'' further out does not lead to a detonation.
This is likely due to the lack of precompression through the He detonation that the CO material would naturally undergo.
In a more complete setup even more off-center CO detonations may be possible (see, e.g., \citealt{gronow2021a} for various double detonation mechanisms); employing larger ignition bubbles can also force an ignition at such densities.
As we show in Section~\ref{sec:results}, our toy models are nonetheless a good representation of more complex double detonation scenarios.

After placing the ignition spot, we let the simulation evolve until the ejecta had securely reached homologous expansion, i.e., our last simulation snapshot is taken $100$\,s after the ignition.
Because nuclear burning ceases after less than $1$\,s, unbinding all material, the ejecta will be in homologous expansion (to the percent level) by $100$\,s \citep{fink2010a,wilk2018a}.

\subsection{Nucleosynthetic post-processing}\label{sec:nucleo}
To allow for a more precise analysis of the nucleosynthetic yields and to generate the inputs for a detailed radiative transfer calculation, we post-processed the nucleosynthesis of the \textsc{Arepo} simulation.
Here, we advected $10^6$ Lagrangian tracer particles in the hydrodynamic flow and recorded the thermodynamic conditions along their trajectories.
These trajectories were then used as input for a much larger, $384$ isotope, nuclear reaction network \citep{seitenzahl2010a,pakmor2012a,seitenzahl2017a}, assuming solar metallicity \citep{asplund2009a}, to obtain the detailed isotopic composition of the ejecta.

\subsection{Radiative transfer}
After the ejecta had reached homologous expansion, their density structure alongside the isotopic composition were mapped to a uniform Cartesian grid that serves as input to the three-dimensional, time-dependent Monte Carlo RT code \textsc{Artis} \citep{sim2007a,kromer2009a,bulla2015a,shingles2020a,sim2024a}, which was built on the methods of \citet{lucy2002a,lucy2003a,lucy2005a}.
Here, we used the approximate non-LTE treatment described by \citet{kromer2009a}.
Specifically, the photoionization rates were determined from the simulation and used to estimate the ionization state.
The excitation states were evaluated through Boltzmann distributions.
Additionally, \textsc{Artis} uses a gray approximation for optically thick cells.
Our radiative transfer calculations use around $8.3\times10^6$ atomic lines with $1.3\times10^5$ levels ($3.6\times10^4$ photoionization transitions) including $142$ ions as input data, based on the \textsc{big\_gf-4} dataset of \citet{kurucz1995a,kurucz2006a}, similar to the one described by \cite{kromer2009a}.

We emphasize that in particular the near-infrared part of the optical spectra will depend heavily on these inputs, especially with respect to the emergence of the second maximum in, for example, the I band \citep{kromer2009a}.
In general, features such as this will heavily depend on the details of the included microphysics \citep{kasen2006a,kromer2009a,jack2015a} and the precise reproduction of observations will require the inclusion of full non-LTE physics (see, e.g., \citealt{blondin2022a,collins2025a} for a study on the timing and strength of the secondary I band maximum or \citealt{shen2021a} for the necessity for a non-LTE solution to reproduce accurate decline rates).
Consequently, we do not expect our results to fully recover absolute quantities that are sensitive to these inputs, such as $\Delta \mathrm{m}_{15}(\mathrm{B})$.
However, since we are mainly interested in a comparison with studies employing similar approximations (specifically \citealt{gronow2021a,pollin2024a}), as well as a differential comparison between different explosions of the same WDs but different ignition points, we trade precision for reduced computational costs required for this parameter study.
Although we expect our models to show the general trends, given recent non-LTE simulations \citep{shen2021a,boos2024b,collins2025a} the exact quantitative values are likely to change.
However, given full 3D non-LTE simulations have not yet been carried out in this context and the expense of non-LTE calculations, a 3D non-LTE RT parameter study is not currently computationally feasible.
Our findings should therefore be reexamined when such a parameter study becomes possible.

We mapped the nucleosynthetically post-processed ejecta onto a $50^3$ Cartesian grid and derived synthetic observables by following $10^8$ energy packets for 100 logarithmically spaced timesteps from $2$ to $60$ days after the explosion, whereby we assume homologous expansion for the entire duration.
After the simulation had finished, we extracted $100$ line of sight-dependent synthetic observables by binning escaping packets into uniformly spaced $\cos \theta$ and $\phi$ solid angle bins.
Here, $\cos \theta$ runs along the $z$ axis and $\phi$ projects the result onto the $x$-$y$ plane.

\section{Results}\label{sec:results}

\subsection{Ejecta composition and structure}\label{sec:res_hydro}
\begin{figure*}
    \centering 
    \includegraphics[trim={0 0 0 1.8cm},clip]{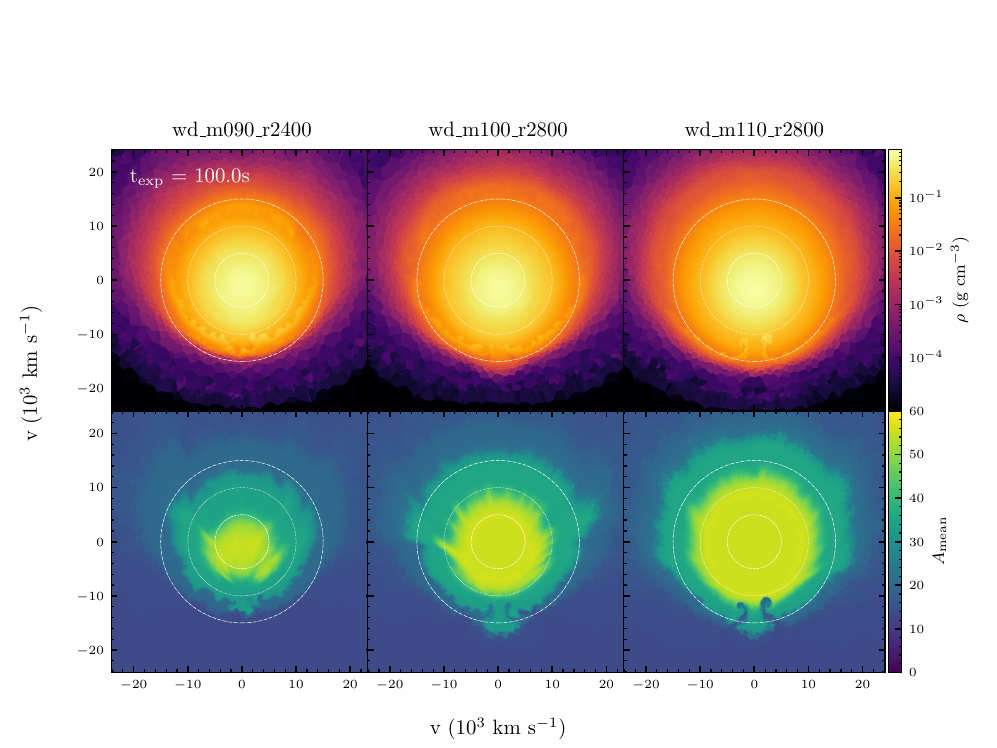}
    \caption{Illustration of an ejecta slice at homology of our most asymmetrically ignited simulations. The slice is taken in the $x$-$y$ plane, the ejecta are roughly rotational symmetric along the $x$ axis. The upper row illustrates the density profile, whereas the lower row shows the mean atomic mass number, calculated from the species used in the in situ network included in \textsc{Arepo}. The circles indicate spherical contours at $5$, $10$, and $15\times10^3\,\mathrm{km}\,\mathrm{s}^{-1}$, indicating the asphericity of the ejecta.}
    \label{fig:ejecta}
\end{figure*}
After reaching homologous expansion, the ejecta structure is illustrated in Figure~\ref{fig:ejecta} for the different models.
Here, we only show the most asymmetric models of each series as the central ignition produces effectively spherical ejecta and the structure transitions to the illustrated ones in a straightforward fashion.
The ejecta structure itself shows a core region that is almost exclusively composed of $^{56}$Ni, surrounded by a shell of IGEs and IMEs.
Here, $^{56}$Ni can be found at comparably low velocities, whereas the IGEs other than $^{56}$Ni and IMEs show increasingly higher velocities.
This is consistent with other models of double detonations (see, e.g., \citealt[figure~2]{pakmor2022a} or \citealt[figure~1]{collins2023a}).
However, our simulations lack the double-ringed structure that gets imprinted from the initial He-detonation (as can be seen in, e.g., \citealt[figure~1]{collins2022a} or \citealt[figure~1]{pollin2024a}), which we are not simulating in this parameter study.

\begin{figure*}
    \centering 
    \includegraphics[trim={0 0 0 1.8cm},clip]{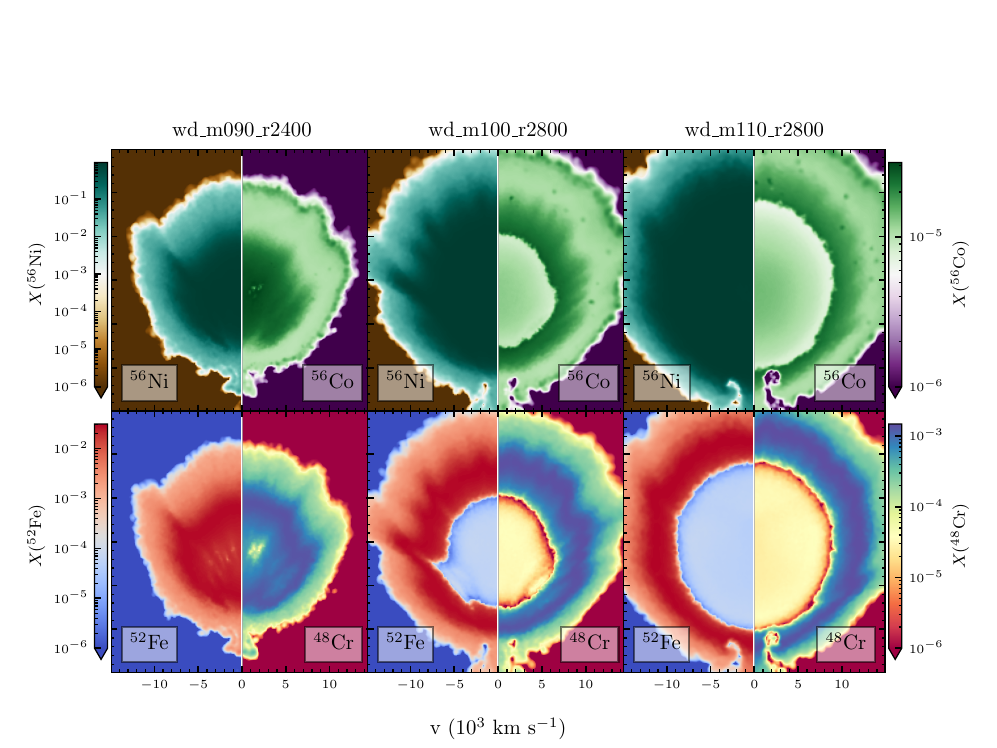}
    \caption{Depiction of radioactives $100\,\mathrm{s}$ post-explosion for our most asymmetrically ignited modes. The slice is taken in the $x$-$y$ plane. Here we illustrate the mapped normalized abundance fractions obtained from the nucleosynthetic postprocessing, as mapped to the radiative transfer simulation. Note that here we show a higher resolution than actually mapped to \textsc{Artis}.}
    \label{fig:radioactives}
\end{figure*}

In Figure~\ref{fig:radioactives} we also show the distribution of some key radioactives for our most asymmetrically ignited models, obtained from the nucleosynthetic post-processing.
Here, it can be seen that $^{56}$Ni is only produced in significant quantities in the central region of the explosion, where the flame burns at the highest densities.
For wd\_m090, the production of $^{56}$Co also follows this region, although the initial formation of a ring-shaped structure is visible in the center.
In the wd\_m100 and wd\_m110 models however, a second $^{56}$Co maximum also appears as a ring-like structure around the central $^{56}$Ni producing region, as well as around the edge of the ejecta.
The lighter $^{52}$Fe and $^{48}$Cr isotopes are located in the burning region at the edge of the central $^{56}$Ni burning region, coinciding with the region where IGEs are produced.
For the wd\_m090 model, these isotopes are produced rather centrally because of the lower burning density, and the ring-like structure can only be seen vaguely in the center.
The distribution of radioactive elements will inevitably influence the synthetic observables, particularly in the early epochs when they release their energy during decay in case of the shorter-lived isotopes.
This will be examined in more detail in Section~\ref{sec:res_rt}.

Detailed isotopic yields obtained from the nucleosynthetic post-processing for all models can be found in Tables~\ref{tab:abundances_090} - \ref{tab:abundances_110}, along with the respective explosion energy.
Overall, the abundances as well as the released energy are in the range of what is expected from WDs of the respective initial mass (see, e.g., \citealt{sim2010a} or \citealt{gronow2021a} for comparable parameter ranges).
There are, however, some anomalies, which we discuss in the following.

First, for the wd\_m090 models, we find that they produce very little $^4$He, whereas the models of \citet{gronow2021a} produce substantial amounts (on the order of $1\times10^{-3}\,\msun$) in their core detonation.
This again comes down to the lack of pre-compression from the He detonation.
In this case it becomes especially important since the wd\_m090 has a considerably lower density than the other two model series, so it barely produces any IGEs at all.
Here, we do not reach temperatures high enough for efficient photodisintegration which would otherwise produce some $^4$He and therefore no $^4$He is produced, at least not in the same quantities as in the models of \citet{gronow2021a}.

Second, when comparing our abundances to the ones found by \citet{gronow2021a}, we find that our models produce somewhat less of key nucleids such as $^{56}$Ni overall for WDs of the same mass.
We argue that this is the result of the missing compression caused by the initial He detonation which we do not simulate in this parameter study. If this compression was included in our model, it would cause our WD to burn at a slightly higher density.

Third, we observe that our off-center ignited models produce more $^{56}$Ni (and in general IGEs) than our centrally ignited counterparts.
For the off-center model, the burning results in slightly higher abundances of $^{56}$Ni in the hemisphere of the ignition since the detonation is somewhat stronger (as an result of the artificial ignition) than in the centrally ignited case, where the detonation has quieted down in comparison.
Additionally, the detonation moves along different gradients, changing the burning dynamics; the centrally ignited model only burns down a density gradient.
Consequently, off-center models produce less $^{56}$Ni in the hemisphere opposite the ignition spot compared to the central ignition, since the detonation has weakened by burning up the density gradient in the center.
However, we emphasize that this effect is only a subtle effect on the percent level, as can be seen in the tables in Appendix~\ref{app:abund}, and is predominantly an artifact of the manual ignition.
Here, the wd\_m110 series exhibits a $2.4$\% increase between the r0 and r2800 model, and the wd\_m100 only a $3.3$\% increase between the same radii.
In case of the wd\_m090 models, this difference is far more substantial, it shows a $46.1$\% increase in the produced $^{56}$Ni mass between the r0 and the r2400 model.
As in the other two series, the increased $^{56}$Ni production for off-center ignition is one factor contributing to this discrepancy, and due to the lower density to an even stronger degree.
In case of the wd\_m090 series, it is not only the off-center models over-producing $^{56}$Ni, but rather the central models under-producing it, which is a result of the artificial ignition.
As in the previous two points, our models lack the pre-compression from the He-detonation, even more so for the wd\_m090 models.
After artificially igniting these, the initial detonation is synthetically powerful and releases too much energy.
As a consequence, the detonation moves faster and the material behind the shock expands too quickly, reducing its density, and thus its subsequent energy output.
If we instead ignite the same wd\_m090\_r0 model with a tenth of the original mass, that is, only $1\times10^{-5}\,\msun$, the initial detonation is less powerful and does not undergo the same shock weakening, at least not to the same extent.
As a result, the shock causes slightly higher compression of the burning material, which in turn causes an increased energy release.
This process takes place within the first few $0.01\,\mathrm{s}$, but compounds to a substantial difference in burn density during the core burning phase and beyond\footnote{We note that this is a rather phenomenological description of the burning behavior for different artificial ignitions. A more fundamental description of this process is outside the scope of this work and will be addressed elsewhere.}.
Here, reducing the size of the ignition spot by one tenth leads to an approximately $25\,\%$ increase in $^{56}$Ni production, which is more in line with the overall $^{56}$Ni production expected from a detonation of a $0.9\,\msun$ WD\footnote{Here we use the results of, e.g., \citet{sim2010a,gronow2021a} as a reference point. It is beyond the scope of this work to say which result is objectively more correct as each of these simulations in turn has its own systematic uncertainties related to the ignition.}.

This large difference is also noticeable in the peak magnitudes of the synthetic light curves (see Section~\ref{sec:res_rt}), but we have no reason to suspect that this fundamentally affects the viewing angle distribution for the synthetic observables.
In order to keep the ignition consistent throughout this study we use the $1\times10^{-4}\,\msun$ ignition spots for the wd\_m090 models, but we note that these models would require a more sophisticated ignition procedure to account, for example, for the missing He detonation.
In general, one could fine-tune the ignition, but since it does not affect the main point of our study, i.e., line of sight effects, we refrain from doing so.

\subsection{Synthetic observables} \label{sec:res_rt}
In this section we present synthetic observables, with a focus on the differential effects due to line of sight variations in the models, rather than on a comparison of, for example, individual light curves to observations.
Such comparisons have already been extensively made in previous work \citep{shen2010a,blondin2017a,shen2021a,shen2021b,collins2022a,pollin2024a,collins2025a} for similar initial masses.
As we do not expect a better match to observational data from our simplified models compared to more sophisticated models presented in previous work, we refrain from an extensive examination of, for example, the light curve or individual spectral features and refer to previous work.

Overall, we are mainly interested in the viewing angle dispersion of our models, and consequently will focus on this aspect; reproducing the overall trend of the Phillips relation requires more sophisticated explosion and RT simulations in any case.
The viewing angle scatter is not connected to the absolute value of $\Delta \mathrm{m}_{15}(\mathrm{B})$, but is rather a relative effect as it is determined mainly through the asymmetry in the ejecta distribution after CO detonation; non-LTE effects would offset them all by roughly the same amount.
Currently, we have no evidence suggesting that the viewing-angle scatter would change substantially after the inclusion of non-LTE effects.
Recent work of \citet[figure 10]{boos2024b} suggests that although there is a slight increase in viewing angle dispersion from LTE to non-LTE RT simulations, it would be a subdominant effect in our argument.
Nevertheless, once multidimensional non-LTE RT photospheric phase simulations are computationally achievable for the set of models discussed in this study, a reassessment will be necessary.

In several places, we compare our results to those of \citet{pollin2024a}.
Their simulations were run with an atomic dataset with fewer lines and therefore their light curves show, for example, slower decline rates.
However, this is only a systematic offset and does not affect the overall viewing angle distribution.
In Appendix~\ref{app:big_gf} we illustrate this systematic offset in decline rate for selected quantities for their \textsc{OneExpl} simulation.

\subsubsection{Light curves}\label{sec:res_lc}

\begin{figure*}
    \centering
    \includegraphics{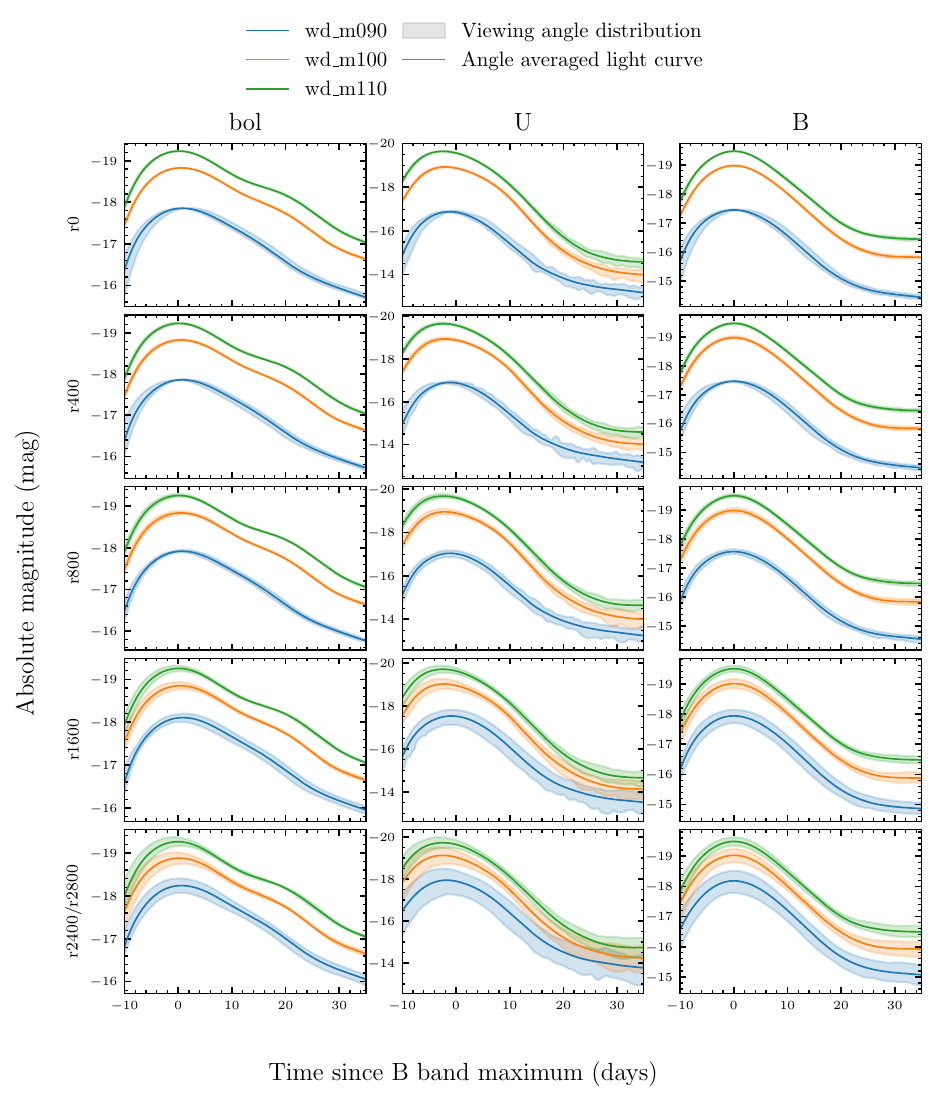}
    \caption{Bolometric, U and B band light curves of our models. Here, the solid lines show the angle averaged light curves, whereas the shaded regions indicate the variability for different viewing angles; that is, the minimum and maximum of each light curve. The jagged features in some of the viewing angle distributions are due to noise in the individual viewing angle light curves.}
    \label{fig:lightcurves1}
\end{figure*}

\begin{table*}
    \centering 
    \caption{Noise floor estimated on the r0 model of each series.}\label{tab:scatter_quants}
    \begingroup
	\setlength{\tabcolsep}{6pt}
	\renewcommand{\arraystretch}{1.25} 
    \begin{tabular}{c | c c c}
    \hline\hline
    Model & $\mathrm{M}_\mathrm{B,max}$\tablefootmark{a} (mag) & $t_{\mathrm{M}_\mathrm{B,max}}$\tablefootmark{a} (days) & $\Delta \mathrm{m}_{15} (\mathrm{B})$\tablefootmark{a} (mag)\\
    \hline
    wd\_m090\_r0 & $-17.45 \pm 0.01\,\,[0.05]$ & $17.62 \pm 0.30\,\,[2.22]$ & $1.70 \pm 0.05\,\,[0.33]$ \\
    wd\_m090\_r400 & $-17.47 \pm 0.02\,\,[0.07]$ & $17.72 \pm 0.32\,\,[2.22]$ & $1.72 \pm 0.05\,\,[0.37]$ \\
    wd\_m090\_r800 & $-17.57 \pm 0.05\,\,[0.17]$ & $17.71 \pm 0.29\,\,[1.33]$ & $1.73 \pm 0.04\,\,[0.20]$ \\
    wd\_m090\_r1600 & $-17.93 \pm 0.15\,\,[0.44]$ & $17.39 \pm 0.44\,\,[2.07]$ & $1.74 \pm 0.04\,\,[0.22]$ \\
    wd\_m090\_r2400 & $-18.16 \pm 0.24\,\,[0.74]$ & $17.44 \pm 0.40\,\,[1.73]$ & $1.80 \pm 0.08\,\,[0.34]$ \\
    \hline
    wd\_m100\_r0 & $-18.98 \pm 0.01\,\,[0.03]$ & $17.17 \pm 0.07\,\,[0.35]$ & $1.78 \pm 0.02\,\,[0.08]$ \\
    wd\_m100\_r400 & $-18.98 \pm 0.03\,\,[0.09]$ & $17.18 \pm 0.08\,\,[0.35]$ & $1.78 \pm 0.02\,\,[0.10]$ \\
    wd\_m100\_r800 & $-18.99 \pm 0.05\,\,[0.17]$ & $17.18 \pm 0.10\,\,[0.44]$ & $1.79 \pm 0.04\,\,[0.15]$ \\
    wd\_m100\_r1600 & $-19.00 \pm 0.10\,\,[0.31]$ & $17.23 \pm 0.13\,\,[0.59]$ & $1.80 \pm 0.06\,\,[0.20]$ \\
    wd\_m100\_r2800 & $-19.02 \pm 0.15\,\,[0.44]$ & $17.35 \pm 0.26\,\,[1.04]$ & $1.83 \pm 0.08\,\,[0.27]$ \\
    \hline
    wd\_m110\_r0 & $-19.47 \pm 0.01\,\,[0.03]$ & $16.40 \pm 0.08\,\,[0.44]$ & $1.75 \pm 0.02\,\,[0.08]$ \\
    wd\_m110\_r400 & $-19.47 \pm 0.02\,\,[0.07]$ & $16.36 \pm 0.09\,\,[0.39]$ & $1.75 \pm 0.02\,\,[0.11]$ \\
    wd\_m110\_r800 & $-19.49 \pm 0.04\,\,[0.11]$ & $16.25 \pm 0.13\,\,[0.59]$ & $1.76 \pm 0.03\,\,[0.12]$ \\
    wd\_m110\_r1600 & $-19.49 \pm 0.07\,\,[0.21]$ & $16.27 \pm 0.25\,\,[0.99]$ & $1.77 \pm 0.04\,\,[0.17]$ \\
    wd\_m110\_r2800 & $-19.48 \pm 0.10\,\,[0.28]$ & $16.39 \pm 0.41\,\,[1.48]$ & $1.79 \pm 0.06\,\,[0.20]$ \\
    \hline
    \end{tabular}
    \endgroup
    \tablefoot{
    \tablefoottext{a}{The uncertainties refer to the standard deviation around the specified value, whereas the values in brackets indicate the total width of the viewing angle distribution.}
    }
\end{table*}

Figures~\ref{fig:lightcurves1} and \ref{fig:lightcurves2} illustrate the viewing angle variability of the light curves as well as the angle-averaged light curves for some selected radii.
We exclude the r100 and r200 models here for brevity; they do not show any significant increase in viewing angle scatter compared to the r0 model series.
In these figures, the individual light curves have been centered around their individual B band maximum and not the B band maximum in the angle-averaged case.

With respect to the overall light curve evolution, we note that, similar to for example \cite{collins2022a}, our light curves decline too fast after around two weeks post-peak luminosity likely because of the lack of a full non-LTE treatment in our RT simulations; both the decline rate as well as the appearance of the NIR secondary maximum are affected by this.
The inclusion of non-LTE effects seems to remedy this discrepancy (see, e.g., \citealt{collins2025a}).

Regarding the viewing angle dependence, several things stand out.
First, we note the nonzero viewing angle dispersion in the r0 case; in an ideal scenario this series should be equivalent to an angle-averaged simulation with zero dispersion.
However, due to numerical noise in the explosion simulation (e.g., discretization errors of the otherwise perfectly spherically symmetric ignition spot) the explosion is not perfectly symmetric and shows small-scale perturbations.
Additionally, the RT calculation is also affected by noise, in particular the individual viewing-angle bins.
In the more centrally ignited models of the wd\_m090 series and the wd\_m090\_r0 model in particular, this problem is amplified by the overall lower $^{56}$Ni mass and the issues connected to the artificial ignition, see Section~\ref{sec:res_hydro}.
Therefore, the larger dispersion of those models should be considered unphysical.
This problem is difficult to overcome without spending substantially more computational resources, i.e., increasing the packet number, which is out of scope for the current project.
As a result, there is some intrinsic variability in the peak magnitudes and time evolution of the individual light curves, leading to nonzero variability even in the centrally ignited models.
We quantify this noise floor through the standard deviation of both the peak B band magnitude and the epoch of maximum B band brightness. The corresponding values can be found in Table~\ref{tab:scatter_quants}; that is, the standard deviation on the respective values for the r0 models.
Overall, we find this noise floor to be a subdominant effect in our following analysis.

\begin{figure*}
    \centering
    \includegraphics{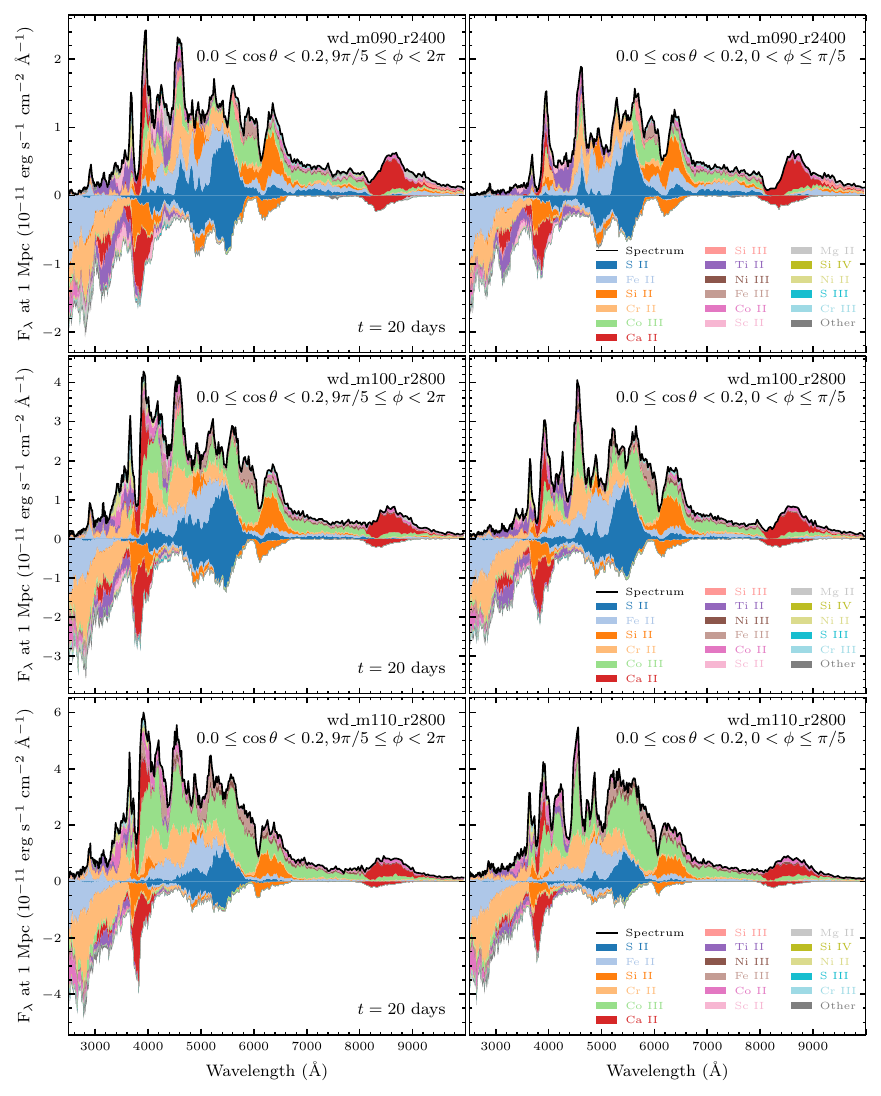}
    \caption{Spectral energy density for our most asymmetrically ignited models. Here we show two opposing lines of sight $20$\,days after the explosions. The left column corresponds roughly to the north pole of Figure~\ref{fig:ejecta}, whereas the right column point toward the south pole. In each panel, the black line describes the overall synthetic flux, while colored layers illustrate the contributions of individual ions to the overall emitted and absorbed flux. Here, the largest difference in flux can be seen in the blue wavelengths below approximately $4500$\,\AA{}, caused by line blanketing of certain IGEs.}
    \label{fig:spec_va}
\end{figure*}

Second, looking at the dispersion of the more off-center ignitions, we see that only after $400$\,km, in the r800 series, a substantial increase in the light curve variability can be observed with the r2400/r2800 models showing quite a significant line of sight effects (see Table~\ref{tab:scatter_quants}).
Generally, the connection between the viewing angle distribution and the ejecta asymmetry has already been examined previously (see, e.g., \citealt{kromer2010a,sim2012a,gronow2020a,collins2022a,boos2024a, pollin2024a}).
In short, there are two effects at play here:
On the one hand, the central $^{56}$Ni distribution is somewhat asymmetric (see Figure~\ref{fig:radioactives}), causing the side that is closer to the surface to rise faster in brightness and, in general, to higher peak magnitudes.
On the other hand, the opposite side will be rich in lighter IGEs (such as Ti or Cr), which lead to substantial line blanketing in this direction, further reducing its luminosity, especially in the blue bands.
This is, generally speaking, the leading cause for the viewing angle dependence of the light curves.
Another effect well reproduced by our models is the reduced viewing angle dispersion in the redder bands, examined in more detail by \cite{pollin2024a}.
The typical opacity in the red is lower, meaning that the emission in these bands, often driven by fluorescence from bluer wavelengths, is less affected by the overlying opacity distribution, and thus less angle-dependent.
We illustrate the effect for two opposing lines of sight in Figure~\ref{fig:spec_va}.
There, the impact of the line blanketing on the flux in the region below $4500$\,\AA{} can be clearly seen.

Third, we note that the wd\_m090 series shows the overall largest viewing angle dispersion in terms of magnitude difference.
This is caused by the overall asymmetry of all key elements and the radioactives being located in the central region at higher densities, compared to a ring-like structure as in the more massive series (see Figures~\ref{fig:ejecta} and \ref{fig:radioactives}).
Additionally, the wd\_m090 series produces more line blanketing elements relative to its $^{56}$Ni production, causing its effect to be more pronounced for the given flux intensity and, as such, leading to larger viewing angle variability.

\subsubsection{Color evolution}\label{sec:color_evo}

Closely related to the light curves, we next examine the color evolution of our models, illustrated in Figure~\ref{fig:color_evolution}.
Here, the first thing that stands out is that our models show a general evolution from bluest around peak to redder at later times, which is reminiscent of what is observed (see, e.g., \citealt{hicken2009a}).
The second important detail is that, as one would naively expect due to the correlation between the scatter in different passbands, the viewing angle scatter of the colors is less than the scatter of the individual light curves.
Therefore, the variation in the colors with different lines of sight are relatively small in most cases.
Models including a He-shell historically struggle with being too red compared to observations (see, e.g., \citealt{kromer2010a,woosley2011a,polin2019a,gronow2020a,shen2021b,collins2022a}) because the elements produced in the He detonation have a strong line blanketing effect.
The problem is somewhat remedied when a minimal He mass is used \citep{townsley2019a,shen2021b}; the inclusion of non-LTE effects further improves the color evolution \citep{collins2025a}.

Strongly linked to this line blanketing effect and the second point made in the preceding Section~\ref{sec:res_lc}, our red colors show rather little viewing angle dispersion, an effect that is strongly connected to the asymmetric distribution of elements that cause line blanketing.
Given the absence of heavy elements produced in the He detonation, there is less line blanketing at these epochs and consequently the spectrum will appear bluer; the wd\_m090 is more dominated by line blanketing at these epochs due, for example, to higher relative Cr production.
However, there is another important effect at play:
The lower luminosity means that the fainter wd\_m090 models have a lower degree of ionization.
The lower ions of the IGEs (particularly Ti, Cr, Fe, Co, and Ni) are generally better at blanketing the blue wavelengths than the higher ions.
So while the overall larger amount of these elements contributes to the asymmetry, the larger fraction of singly or doubly ionized states makes line blanketing more effective than the more luminous models exhibiting higher ionization states.

This is consistent with the findings of \citet{pollin2024a} for their \textsc{NoHe} models.
More importantly, we compare these color curves and their viewing angle dependence with the results of \citet{pollin2024a}, illustrated in the last row of Figure~\ref{fig:color_evolution}.
Examining the overall magnitude of the viewing angle dispersion, we find that our more massive models reproduce the dispersion of the \textsc{3DOneExplNoHe} model of \cite{pollin2024a}.
Focusing on our series of wd\_m100 models, which is closest in mass to their exploding WD mass, we observe a \mbox{B-V} viewing angle dispersion of around $0.3$\,mag at around $15$\,days, consistent with the around $0.2$\,mag dispersion of their \textsc{3DOneExplNoHe} model.
In contrast, our viewing angle variability is inconsistent with that found in their \textsc{3DOneExpl} model ($\approx0.6$\,mag) even for our most asymmetric model series.
We also note that only our wd\_m100\_r1600 and wd\_m100\_m2800 are qualitatively consistent with their \textsc{3DOneExplNoHe} models in terms of viewing angle dispersion, while more central models show substantially less viewing angle variability.

\subsubsection{Peak magnitudes and the Phillips relation}\label{sec:peakmag}

As summarized in Section~\ref{sec:intro}, the main motivation for systematically investigating off-center detonations is the fact that even rather sophisticated models struggle to reproduce the scatter around the Phillips relation through their viewing angle dispersion (see \citealt{gronow2021a,collins2022a,pollin2024a,boos2024a}) with their scatter being substantially outside the range of scatter of observed SNe.
In the following, we explore our systematically offset ignition models to examine if off-center CO-detonations are consistent with astronomical data.

\begin{figure}
    \centering
    \includegraphics{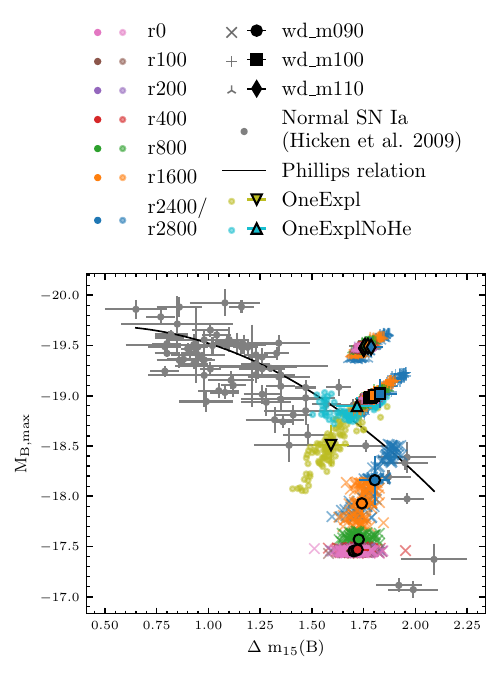}
    \caption{Illustration of the peak B band magnitudes and decline rates of our models in the context of the Phillips relation. Here the translucent glyphs represent the individual viewing angles while the symbols outlined in black indicate the mean values and the connected standard deviations. The peak B band magnitude is calculated for the peak of each line of sight individually. We also show observational data of \citet{hicken2009a}. Absolute magnitudes for the observational data have been computed assuming $H_0 = 70.2$\,km\,s$^{-1}$\,Mpc$^{-1}$. Objects with a distance modulus of $\mu < 33$ have been excluded. The \textsc{OneExpl} and \textsc{OneExplNoHe} data is the same as the one presented in \citet{pollin2024a}.}
    \label{fig:phillips}
\end{figure}

First, we look at the Phillips relation; that is, M$_\mathrm{B,max}$ versus $\Delta \mathrm{m}_{15}(\mathrm{B})$, illustrated in Figure~\ref{fig:phillips}, which also includes observational data from \cite{hicken2009a}.
It is immediately apparent that our models face significant challenges in accurately replicating the Phillips relation, specifically the correct $\Delta \mathrm{m}_{15}(\mathrm{B})$ values\footnote{The wd\_m090 series also struggles in reproducing the correct peak magnitudes in case of the more centrally ignited models. This is, as explained in Section~\ref{sec:res_hydro} an artifact of the ignition.}.
This is not entirely unexpected, as modeling the decline rate accurately depends heavily on the microphysics included, in particular non-LTE effects.
\citet{shen2021a} recently demonstrated that the inclusion of non-LTE effects results in a rather good match to the Phillips relation for varying WD masses (see also \citealt{blondin2017a}).
As elaborated in the preamble to Section~\ref{sec:res_rt}, we are mainly interested in the viewing angle dispersion, and consequently will focus on this aspect; reproducing the overall trend of the Phillips relation requires more sophisticated explosion and RT simulations in any case.
Setting aside the systematic offset in absolute values, we see that the viewing-angle dispersion predicted by our models, even those with the most off-center ignition, is well within the scale of the observed scatter about the mean relation.
We further compare our models with those of \citet{pollin2024a}\footnote{We remind the reader that \citet{pollin2024a} use a smaller atomic dataset and thus obtain smaller $\Delta \mathrm{m}_{15}(\mathrm{B)}$ values. This is, however, only a systematic offset and does not substantially affect the viewing angle distribution. See Figure~\ref{fig:phillips_big}.}.
As in the previous section, we find good agreement between our wd\_m100 model series and their \textsc{OneExplNoHe} synthetic observables, both in terms of peak magnitude as well as in the decline rate (when comparing to our most asymmetrically ignited model).
The only noticeable deviation is the tail on the slow end of the \textsc{OneExplNoHe} data; this tail is most likely due to the dynamical impact of the secondary WD on the He ejecta structure, see \citet[figure 2]{pakmor2022a}.
This suggests that our toy models are reasonably representative of a more sophisticated setup in terms of the effect of the explosion dynamics on the observables; for the \textsc{OneExplNoHe} model only the He ash itself was removed, but not the effect of the He detonation on the CO fuel.
Although \textsc{OneExplNoHe} is similar to our models and is in agreement with the scatter of the observational data, the \textsc{OneExpl} model shows a substantially increased viewing angle scatter.
As stated previously and is described in greater detail by \citet{pollin2024a}, this is connected to the impact of the He ash on the spectral flux, in particular in the blue wavelengths through line blanketing.
The \textsc{OneExpl} model already exceeds the observed scatter around the Phillips relation; the \textsc{TwoExpl} scenario\footnote{Here, the name \textsc{TwoExpl} refers to a scenario where both the primary and secondary WD explode instead of only the primary WD in case of the \textsc{OneExpl} scenario. The naming scheme reflects the number of WDs that explode and get destroyed. For more details see \citet{pakmor2022a}.}, not depicted here, is even more extreme in this respect, while our argument still holds for the \textsc{TwoExplNoHe} model, cf. \citet[figure 7]{pollin2024a}. 
While the work of \citet{pollin2024a} already hinted at the fact that the He-shell is the driving force behind the significant viewing angle distribution of double detonations, our systematic study confirms that indeed the core detonation does not lead to a viewing angle dispersion at tension with the observations.
We find that, for reasonable models, the degree of angle variation is on the scale that it may indeed be an important contribution to the observed scatter about the Phillips relation -- i.e., an off-center detonation is not at odds with the observed data, and moreover may help explain why there is scatter in the observed relation.

\begin{figure}
    \centering
    \includegraphics{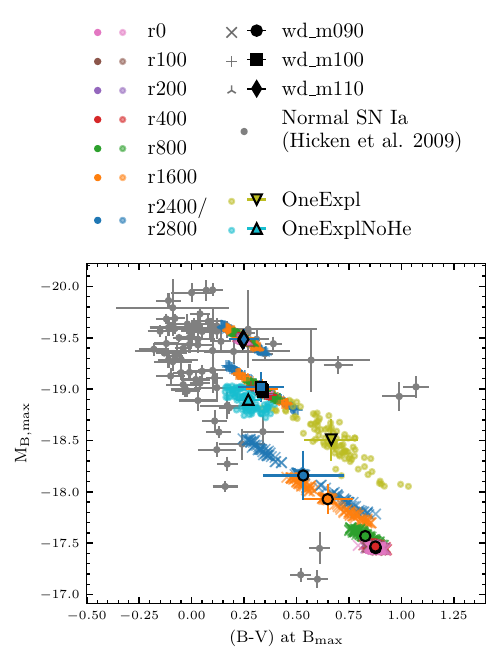}
    \caption{Same as Figure~\ref{fig:phillips}, but here we show the B-V color at peak B band magnitude instead of $\Delta \mathrm{m}_{15}(\mathrm{B})$.}
    \label{fig:mb_color}
\end{figure}

Furthermore, Figure~\ref{fig:mb_color} shows the peak B band magnitude versus the \mbox{B-V} color.
Here, it can be seen that our wd\_m090 model series is too red, due to the aforementioned line blanketing.
This impacts both its mean color as well as an excessive viewing-angle scatter.
Similarly, the \textsc{OneExpl} model of \citet{pollin2024a} also shows a dispersion too large compared to observations.
In contrast, our wd\_m100 and wd\_m110 as well as the \textsc{OneExplNoHe} model fit well with the observations; these models are expected to be slightly too red due to the lack of treatment for non-LTE effects in the RT calculations; see the previous section.

\begin{figure}
    \centering
    \includegraphics{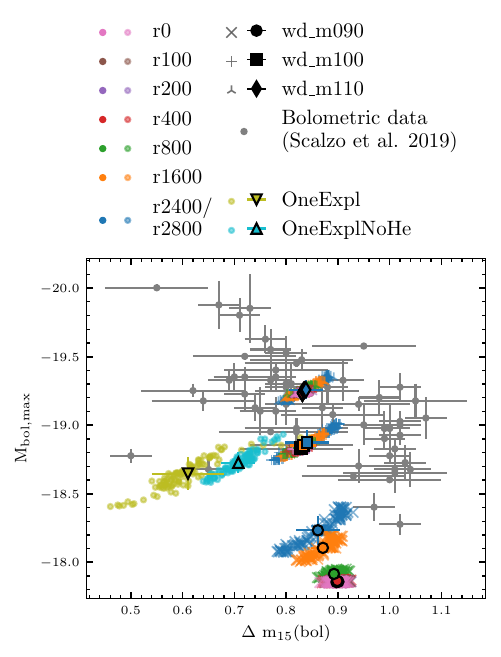}
    \caption{Same as Figure~\ref{fig:phillips}, but here we show bolometric instead of B band quantities. The observational data show is taken from \citet{scalzo2019a}.}
    \label{fig:peakmag_m15}
\end{figure}

Lastly, we also examine the peak bolometric luminosity versus the bolometric decline rate, shown in Figure~\ref{fig:peakmag_m15}, and compare it against figure~6 of \citet{gronow2021a}.
It should be noted that, in general, the level of scatter in the bolometric width-luminosity relation is similar to the one found in the B band Phillips relation.
Here, bolometric quantities have the advantage that they also account for the red bands not considered in the previous discussion.
Furthermore, bolometric light curves are less affected by non-LTE effects \citep{dessart2014a}.
It can be seen that our models and the \textsc{OneExplNoHe} model agree well with the observational data of \citet{scalzo2019a}; again, we focus mainly on the viewing angle dispersion, as reproducing the decline rates is difficult without the inclusion of full non-LTE effects.
In contrast, the He-shell models of \citet{gronow2021a} and, to a lesser extent, the \textsc{OneExpl} model of \citet{pollin2024a}\footnote{See Figure~\ref{fig:peakmag_m15_big} for an illustration of the systematic impact of the atomic dataset.} do not reproduce the observed variability (cf. \citealt[figure 6]{gronow2021a}).
Comparing our results with the results of \citet[figure~19]{boos2024a}, we find that our models show a viewing angle scatter that is lower than that of all of their models\footnote{For example, our wd\_m100 model approximately shows a variation of $0.5\,\mathrm{mag}$ in M$_\mathrm{bol,max}$ and $0.12\,\mathrm{mag}$ in $\Delta\mathrm{m}(\mathrm{bol})$, whereas their $1.00\,\msun$ model shows variations of $0.6\,\mathrm{mag}$ and $0.25\,\mathrm{mag}$, respectively.}, i.e., their two explosion models without a He-shell and their one explosion models including a thin He-shell.
A two-explosion scenario and a He-shell seem to each independently influence the distribution of viewing angles.
This is in line with our previous results, but here the shift to bolometric quantities also shows that this behavior is independent of any particular filter bandpass.

\subsection{Spectral features}\label{sec:spec}
\begin{figure}
    \centering
    \includegraphics{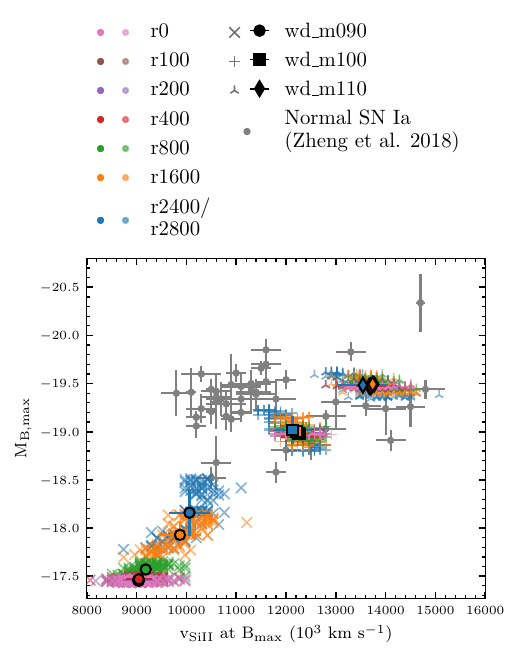}
    \caption{Same as Figure~\ref{fig:phillips}, but here we show $v_\mathrm{\ion{Si}{ii}}$ instead of $\Delta \mathrm{m}_{15}(\mathrm{B})$. The observational data is taken from \citet{zheng2018a}.}
    \label{fig:si_vel}
\end{figure}

Here we examine the line velocity of the \ion{Si}{ii} $6355$\,\AA{} feature, a well-studied property of SN Ia (see, e.g., \citealt{wang2009a,polin2019a}).
Figure~\ref{fig:si_vel} shows the \ion{Si}{ii} line velocities at the peak B band magnitude of each individual line of sight of our models, obtained using the \textsc{Sccala}\footnote{\url{https://github.com/AlexHls/Sccala}} toolkit (Holas et al., in prep.).\footnote{Different methods for obtaining line velocities will inevitably introduce systematic differences.
However, we find that these differences are only on the order of $100$\,km\,s$^{-1}$ and thus subdominant in our analysis.}
It is immediately apparent that our models reproduce the overall scatter rather well, although at a slight tension with the absolute velocity values.
For our wd\_m090 series, the velocities are somewhat too low, with maybe an exception of the wd\_m090\_r2400 model.
Here, it should be mentioned that the sample of \citet{zheng2018a} only contains normal SN~Ia; due to the low luminosities of most of the wd\_m090 series they may not be represented in the sample and thus we do not expect a good agreement.
The wd\_m100 series only partially reproduces normal velocity values ($v_\mathrm{\ion{Si}{ii}} < \num{11800}\,\mathrm{km}\,\mathrm{s}^{-1}$, see \citealt{wang2009a}) with many viewing angle values falling outside that range.
Only the wd\_m110 series matches well with the high-velocity sample of \citet{zheng2018a}, both in terms of velocity and dispersion.
Most importantly, our models show an overall trend where, taken as a whole, the \ion{Si}{ii} line velocity increases with the B band brightness, whereas the data do not really show such a trend: they more or less scatter with a range of velocities that is pretty similar across the range of brightness.

It should be noted that the absolute velocities of our models may not be a good representation of the \ion{Si}{ii} line velocities found in more sophisticated models.
Here, the dynamical interaction between the two detonations and their impact on the ejecta morphology will have a pronounced effect on quantities such as $v_\mathrm{\ion{Si}{ii}}$.
In fact, looking at the results of \citet[figure~18 and 19]{pollin2024a}, we see that their \textsc{OneExplNoHe} model, which so far has been a good match to our wd\_m100 series, is rather far from our results.
Although we can conclude that an off-center CO detonation alone does not produce $v_\mathrm{\ion{Si}{ii}}$ values at tension with observations in terms of viewing angle dispersion, it is hard to draw any deeper conclusions from this, as we lack a preceding He-detonation and its impact on the ejecta dynamics.
A more detailed investigation of this behavior utilizing more complete explosion simulations will be left to future work.

\section{Discussion}\label{sec:discussion}
The main conclusion from these results is that an off-center CO ignition (in a one-explosion scenario), while causing considerable line of sight effects, is unlikely to be the driving force behind the excessive viewing angle scatter observed by, for example, \citet{gronow2021a,collins2022a,boos2024a}.
However, the off-center detonation contributes substantially to the overall viewing angle distribution within the range of observed values.
As demonstrated in Figure~\ref{fig:phillips}, our most off-center models reproduce a non-negligible amount of the observed scatter around the Phillips relation.
Ignoring this effect by, for example, igniting the CO detonation only in the center, will grossly underestimate the viewing angle dependence of the resultant observables.
Therefore, it is paramount to correctly capture the location of the second location in future hydrodynamical explosion simulations to accurately reproduce the observed scatter.
Additionally, it must be considered that the interaction with a potential detonation in the secondary WD (see the \textsc{TwoExpl} models of \citealt{pollin2024a}), further increases the viewing angle dispersion.
In such a case, the maximum off-center ignition location of the primary CO WD may be even more limited than in the case studied here; this will be examined separately in future work.

Our parameter study suggests that the He detonation and resultant ashes are the dominant factors contributing to the excessive viewing angle dispersion in the aforementioned double detonation models of, for example, \citet{gronow2020a,collins2022a,boos2024a}
This is particularly due to the excessively large mass of the He-shell in the respective initial models.
Both our B band and bolometric decline rate investigations (Figures~\ref{fig:phillips} and \ref{fig:peakmag_m15}) show that our models fit well within the observations, at least to the extent we can determine in the framework of this study.
This is supported by the findings of other authors \citep{collins2022a,pollin2024a} that the ashes of a He detonation contain substantial amounts of elements that contribute significant line blanketing and generate viewing angle scatter due to their asymmetric distribution.
Our models, as well as, for example, the \textsc{OneExplNoHe} model of \citet{pollin2024a}, do not contain these ashes.
Although these models show substantial line blanketing in the blue wavelengths, it is within the observationally expected range.
We arrive at the conclusion that the excessive viewing angle scatter is primarily attributed to the overproduction of He detonation ashes, rather than to asymmetries induced by an off-center detonation of the CO core in the double detonation explosion scenario.
Decreasing the mass of the He ashes could potentially mitigate this problem.

Typically, He-shell masses in double detonation simulations range from $0.1\,\msun$ \cite{woosley2011a,neunteufel2016a} down to $0.02\,\msun$ \cite{fink2010a,townsley2019a}.
However, for this mass range \citet{gronow2021a,collins2022a} found that neither of these He-shell masses is able to reproduce the observed scatter around the Phillips relation.
This also includes the model of \citet{pakmor2022a} which used a He-shell of $0.03\,\msun$ (see the RT calculations of \citealt{pollin2024a}).
Our work suggests that these He-shells are too massive to produce ejecta and subsequent synthetic observables compatible with observations, in agreement with findings of other authors (see, e.g., \citealt{polin2019a,shen2021b}).
Lower shell masses may be able to resolve this tension, at least in a one-explosion scenario; the two-explosion scenario seems to introduce additional dynamical effects that increase the scatter even in the absence of a He-shell (see \citealt{boos2024a}).
\citet{shen2014a} showed that He-shell masses as low as $0.01\,\msun$ produce double detonations with ashes that have a composition compatible with observations.
It remains to be seen whether a He-shell of $0.01\,\msun$ is already sufficient to reproduce the observed distribution or if a lower He-shell mass is necessary.

Some other points are also worth considering in the context of our parameter study.
First, our study is only representative of converging shock double detonation mechanisms.
Other mechanisms, such as the edge-lit or the scissor mechanism \citep{gronow2021a} are not within the scope of our model series, as we cannot sensibly ignite CO WDs that far off-center without the compression of the preceding He detonation.
However, the results of \citet{gronow2021a,collins2022a} suggest that these mechanisms do not increase the viewing angle scatter significantly, at least not when including the contributions of the He ashes.
A more detailed investigation of these mechanisms and the \textsc{NoHe} viewing angle scatter will be left to future work.

Second, we point out that quantities such as the \ion{Si}{ii} line velocity should also be investigated to constrain the viewing angle scatter of models.
However, this requires models more complete than ours.
As demonstrated in Figure~\ref{fig:si_vel}, our simulation, although not in direct tension with the observations in terms of viewing angle scatter, does not accurately reproduce the overall velocity distribution expected from the observations, and show a bulk correlation between luminosity and velocity not exhibited by the data.
There remains a point to be made that a light He-shell may not influence the CO detonation ashes to the same extent as the He-shell detonations of current models and thus bring the $v_\mathrm{\ion{Si}{ii}}$ values of such models closer to ours.
Additionally, it is unclear whether a light He-shell can resolve the tension with respect to the bulk correlation between luminosity and velocity found in our models that is in tension with the observational data.

Third, as was previously stated, the lack of a preceding He detonation will impact the CO detonation ash structure, changing its asymmetry.
Still, within the converging shock scenario, we have no evidence that this would impact the resulting observables to an extent that would change any of the points made in this study.

Combining our results with the findings of various other works \citep{townsley2019a,shen2021a,gronow2021a,pakmor2022a,collins2022a,boos2024a,pollin2024a}, we postulate the following interpretation of the Phillips relation and, subsequently, the normal SNe Ia:
The mean trend of the Phillips relation is given by the mass of the exploding WD in the case of a one-explosion scenario or effective mass in case of a two-explosion scenario, as shown by \citet{shen2021a,boos2024a}.
The scatter around this relation is given by the viewing angle dependence of the observations.
Here, the width of this distribution could yield further conclusions about the average He-shell mass of the WD progenitors of double detonations.
This theory lacks explicit confirmation by models, but with advances in numerical modeling, thin He-shell ($\leq 0.01\,\msun$) models may become available and we will be able to test this conjecture directly.
We acknowledge that this hypothesis ignores the potential contribution to the normal SN Ia population through other explosion channels (such as the detonation-to-deflagration transition scenario; see, e.g., \citealt{seitenzahl2013a}) or the influence of the explosion of the secondary WD.
Much more thorough investigation will be necessary to fully confirm or reject this speculation.

\section{Conclusions}\label{sec:conclusion}
In this work, we presented a systematic parameter study of off-center CO detonations in sub-$M_\mathrm{Ch}$ WDs.
We investigated three initial models of $0.9$, $1.0$, and $1.1\,\msun$, each ignited at seven different offset locations.
In Section~\ref{sec:res_hydro}, we examined the resulting ejecta structure and found that our simplified toy models show a similar ejecta structure, both in density and composition, to more sophisticated models such as those of \citet{gronow2021a,pakmor2022a,boos2024a}, albeit without the presence of He ashes.
Here, we emphasized the asymmetrically distributed layer of IMEs and lighter IGEs around a central core of $^{56}$Ni.
In terms of synthetic observables, we focused our investigation on the viewing angle distribution, rather than a thorough reproduction of observed SNe.

This viewing angle dispersion effect was investigated in Section~\ref{sec:peakmag}.
Our results suggest that even our most asymmetrically ignited models are consistent with the degree of scatter in observed SNe Ia, while models including He-shells exceed the observed viewing angle distributions.
A comparison against the \textsc{OneExplNoHe} model of \citet{pollin2024a} shows that our viewing angle distribution is consistent with that of a more complex explosion mechanism, i.e., including the preceding compression of the CO material through a He detonation.

In Sections~\ref{sec:res_lc} and \ref{sec:color_evo}, we showed the light curves and the color evolution of our models.
We found that in the early phases, our models are less red than the models of, for example, \citet{pollin2024a} because we lack the He ashes that would cause additional line blanketing in the blue wavelengths.
Moreover, we observed that the viewing angle dispersion only substantially increases for models ignited at radii above $400$\,km.

The \ion{Si}{ii} velocities, discussed in Section~\ref{sec:spec}, however, do not match those found by \citet{pollin2024a}.
It is not yet clear if the explosion dynamics of a lighter He-shell would bring their $v_\mathrm{\ion{Si}{ii}}$ values closer to ours.
At least in our simplified pictures, our models reproduce the observed \ion{Si}{ii} velocities rather well.

The main result of our work is that an off-center CO detonation can cause substantial line of sight effects well within the range of what is expected from observational data; accurately simulating the CO detonation ignition location will be required by future models to explain the observed scatter around the Phillips relation.
The excessive viewing angle distributions found by many other authors is likely caused by too massive He-shells and the contributions of their ashes to the spectral energy density, in particular the asymmetric distribution of line blanketing elements.
We have no evidence that off-center CO ignition in the case of a converging shock mechanism is the driving force behind the excessive viewing angle distribution.
Hence, future work should focus on minimizing the He-shell mass in double detonation scenarios.
Other double detonation mechanisms, such as edge-lit or scissor mechanisms, should be considered separately.
Moreover, our study does not cover the effects of an off-center CO detonation on nebular phase observables, which we shall address in a separate study.
Future studies should also focus on including non-LTE effects as they reduce the line blanketing from light IGEs and produce colors more in line with normal SNe Ia \citep{collins2025a}.
Since these line blanketing species are relevant to the viewing angle scatter, non-LTE effects should be investigated with future multidimensional non-LTE simulations.

\begin{acknowledgements}
A.H. is a fellow of the International Max Planck Research School for Astronomy and Cosmic Physics at the University of Heidelberg (IMPRS-HD) and acknowledges financial support from IMPRS-HD.

This work received support from the European Research Council (ERC) under the European Union’s Horizon 2020 research and innovation programme under grant agreement No.\ 759253 and 945806, the Klaus Tschira Foundation, and the High Performance and Cloud Computing Group at the Zentrum f{\"u}r Datenverarbeitung of the University of T{\"u}bingen, the state of Baden-W{\"u}rttemberg through bwHPC and the German Research Foundation (DFG) through grant no INST 37/935-1 FUGG.
The authors gratefully acknowledge the Gauss Centre for Supercomputing e.V. (www.gauss-centre.eu) for funding this project by providing computing time through the John von Neumann Institute for Computing (NIC) on the GCS Supercomputer JUWELS at Jülich Supercomputing Centre (JSC).
The authors acknowledge support by the state of Baden-Württemberg through bwHPC
and the German Research Foundation (DFG) through grant INST 35/1597-1 FUGG.
This work was supported by the Deutsche Forschungsgemeinschaft (DFG, German Research Foundation) -- RO 3676/7-1, project number 537700965,
and by the European Union (ERC, ExCEED, project number 101096243). Views and opinions expressed are, however, those of the authors only and do not necessarily reflect those of the European Union or the European Research Council Executive Agency. Neither the European Union nor the granting authority can be held responsible for them.
FPC and SAS, acknowledge funding from STFC grant
ST/X00094X/1. 
JMP acknowledges the support of the Department for Economy (DfE).
This work used the DiRAC Memory Intensive service Cosma8 at Durham University, managed by the Institute for Computational Cosmology on behalf of the STFC DiRAC HPC Facility (www.dirac.ac.uk). The DiRAC service at Durham was funded by BEIS, UKRI and STFC capital funding, Durham University and STFC operations grants. DiRAC is part of the UKRI Digital Research Infrastructure.
This project has received funding from the European Union’s Horizon Europe
research and innovation programme under the Marie Skłodowska-Curie grant
agreement No.~101152610.
\end{acknowledgements}

\bibliographystyle{aa}
\bibliography{references}

\onecolumn
\begin{appendix}
\section{Light curve and color evolution}
\begin{figure}[h!]
    \centering
    \includegraphics{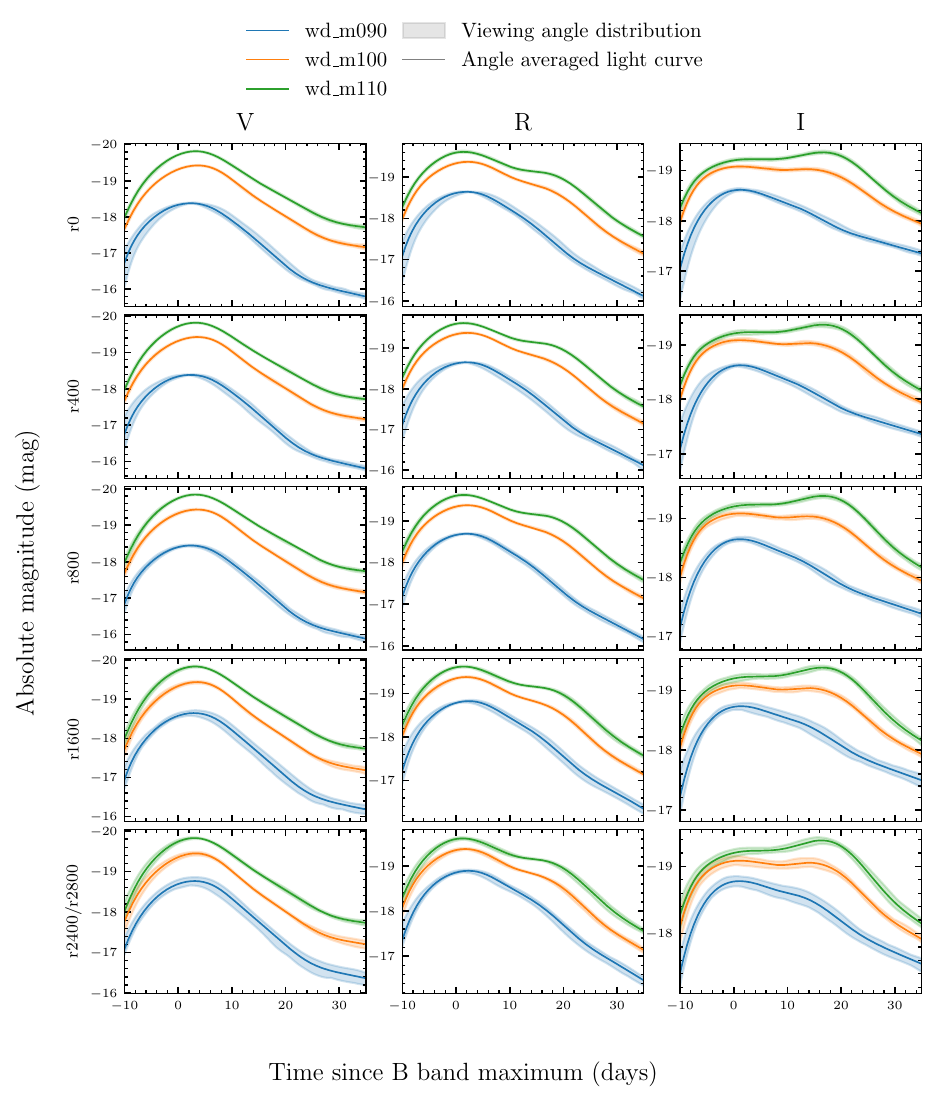}
    \caption{Same as Figure~\ref{fig:lightcurves1}, but for the V, R, and I band light curves.}
    \label{fig:lightcurves2}
\end{figure}

\begin{figure}[h!]
    \centering
    \includegraphics{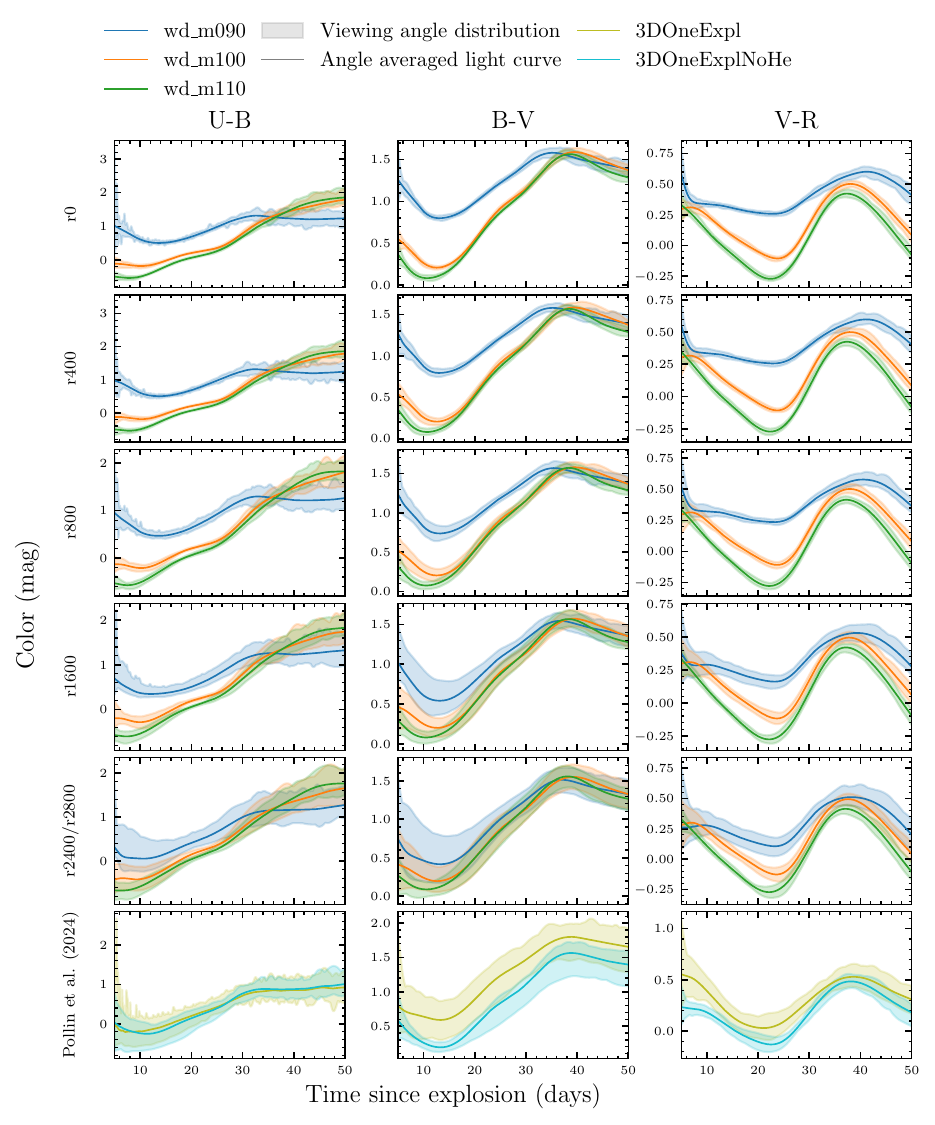}
    \caption{Same as Figure~\ref{fig:lightcurves1}, but here we show the U-B, B-V, and V-R colors. In the last row, we also show the \textsc{OneExpl} and \textsc{OneExplNoHe} model of \citet{pollin2024a} for ease of comparison.}
    \label{fig:color_evolution} 
\end{figure}

This section contains additional light curve and color evolution figures. Figure~\ref{fig:lightcurves2} contains V, R, and I band magnitudes for select models similar to Figure~\ref{fig:lightcurves1}.
Figure~\ref{fig:color_evolution} shows several colors for our models as well as the \textsc{OneExpl} and \textsc{OneExplNoHe} model of \citet{pollin2024a}.
\FloatBarrier

\FloatBarrier

\section{Atomic data comparison}\label{app:big_gf}
\begin{figure}[!hbtp]
    \centering
    \includegraphics{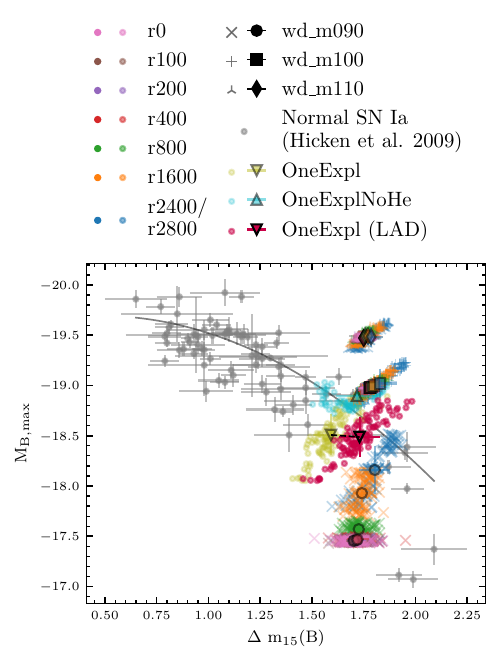}
    \caption{Same as Figure~\ref{fig:phillips}, but here we illustrate the difference that a larger atomic dataset (LAD) makes in terms of decline rate in the case of the \textsc{OneExpl} simulation of \citet{pollin2024a}.}
    \label{fig:phillips_big}
\end{figure}

\begin{figure}[!hbtp]
    \centering
    \includegraphics{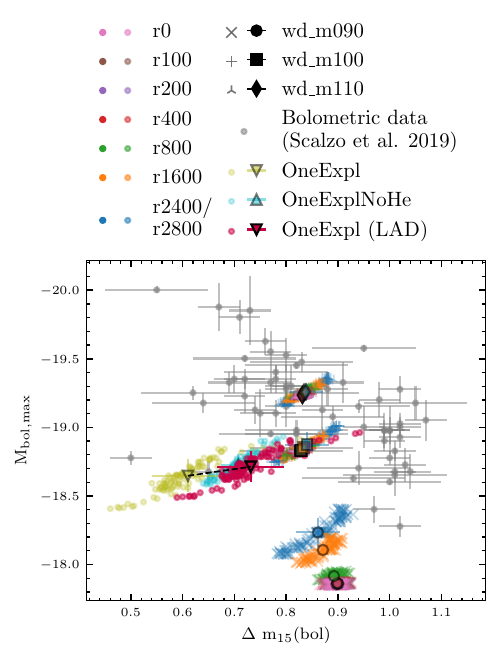}
    \caption{Same as Figure~\ref{fig:peakmag_m15_big}, but here we show the bolometric quantities instead of the B band values.}
    \label{fig:peakmag_m15_big}
\end{figure}

In this section, we show the systematic offset caused by the difference in atomic datasets between our simulations and the simulations of \citet{pollin2024a}, i.e., the \textsc{big\_gf-4} and the \textsc{cd23\_gf-20} datasets described by \citet{kromer2009a}, respectively.
For this purpose, we illustrate a version of their \textsc{OneExpl} model that has been computed using the same \textsc{big\_gf-4} atomic data.
The resulting difference in terms of B band and bolometric decline rate can be seen in Figures~\ref{fig:phillips_big} and \ref{fig:peakmag_m15_big}, respectively.
It can be seen that this leads to a systematic offset of all viewing angles equally and brings the results closer to the decline rates of our models, the wd\_m100 model series in particular.
Consequently, we expect that the \textsc{OneExplNoHe} would match our wd\_m100 model series rather well, both in the decline rate and in the peak magnitude.

\FloatBarrier

\section{Detailed abundance tables}\label{app:abund}
Tables~\ref{tab:abundances_090}, \ref{tab:abundances_100}, and \ref{tab:abundances_110} show the detailed abundances of our wd\_m090, wd\_m100, and wd\_m110 model series, respectively.
The values shown here have been obtained by the nucleosynthetic post-processing described in Section~\ref{sec:nucleo}.
These tables include the energy released in the explosion $\epsilon_\mathrm{expl}$.
In case of the wd\_m090\_r0 model, we also include the abundances that we obtain when igniting this model with an ignition spot containing one tenth of the mass compared to our fiducial ignition setup, i.e., only $10^5\,\msun$.
Here, it can be seen that this greatly improves the production of elements such as $^{56}$Ni and that its observed underproduction is indeed an artifact of artificial ignition.
\begin{table}[h!]
    \centering
    \caption{Detailed abundance table for the models of the wd\_m090 series.}
    \label{tab:abundances_090}
    \begingroup
    \begin{tabular}{l|ccccccc}
    \hline
    {} &\multicolumn{7}{c}{wd\_m090}\\
    Yield ($\msun$) &  r0\tablefootmark{a} &  r100 &  r200 &  r400 &  r800 &  r1600 &  r2400 \\
    \hline\hline
    $^{4}$He & \num{4.35e-11} [\num{1.40e-08}] & \num{2.27e-10} & \num{4.32e-11} & \num{4.35e-11} & \num{8.13e-07} & \num{1.60e-05} & \num{7.19e-05} \\
    $^{12}$C & \num{7.87e-03} [\num{7.73e-03}] & \num{7.86e-03} & \num{7.86e-03} & \num{7.86e-03} & \num{7.88e-03} & \num{8.25e-03} & \num{1.02e-02} \\
    $^{16}$O & \num{1.24e-01} [\num{1.13e-01}] & \num{1.24e-01} & \num{1.24e-01} & \num{1.24e-01} & \num{1.23e-01} & \num{1.17e-01} & \num{1.18e-01} \\
    $^{20}$Ne & \num{3.48e-03} [\num{3.06e-03}] & \num{3.48e-03} & \num{3.48e-03} & \num{3.49e-03} & \num{3.45e-03} & \num{3.52e-03} & \num{3.91e-03} \\
    $^{24}$Mg & \num{1.10e-02} [\num{9.62e-03}] & \num{1.10e-02} & \num{1.10e-02} & \num{1.10e-02} & \num{1.08e-02} & \num{1.03e-02} & \num{1.05e-02} \\
    $^{28}$Si & \num{3.19e-01} [\num{2.92e-01}] & \num{3.19e-01} & \num{3.19e-01} & \num{3.18e-01} & \num{3.13e-01} & \num{2.91e-01} & \num{2.70e-01} \\
    $^{32}$S & \num{1.81e-01} [\num{1.72e-01}] & \num{1.81e-01} & \num{1.81e-01} & \num{1.81e-01} & \num{1.78e-01} & \num{1.67e-01} & \num{1.55e-01} \\
    $^{36}$Ar & \num{3.15e-02} [\num{3.17e-02}] & \num{3.15e-02} & \num{3.15e-02} & \num{3.15e-02} & \num{3.11e-02} & \num{2.96e-02} & \num{2.79e-02} \\
    $^{40}$Ca & \num{2.68e-02} [\num{2.86e-02}] & \num{2.68e-02} & \num{2.68e-02} & \num{2.67e-02} & \num{2.65e-02} & \num{2.58e-02} & \num{2.47e-02} \\
    $^{44}$Ti & \num{1.57e-05} [\num{1.78e-05}] & \num{1.57e-05} & \num{1.57e-05} & \num{1.57e-05} & \num{1.56e-05} & \num{1.59e-05} & \num{1.58e-05} \\
    $^{48}$Cr & \num{3.87e-04} [\num{4.71e-04}] & \num{3.88e-04} & \num{3.88e-04} & \num{3.88e-04} & \num{3.91e-04} & \num{4.16e-04} & \num{4.32e-04} \\
    $^{52}$Fe & \num{7.61e-03} [\num{9.42e-03}] & \num{7.63e-03} & \num{7.63e-03} & \num{7.64e-03} & \num{7.82e-03} & \num{8.73e-03} & \num{9.32e-03} \\
    $^{56}$Ni & \num{1.78e-01} [\num{2.22e-01}] & \num{1.78e-01} & \num{1.79e-01} & \num{1.79e-01} & \num{1.89e-01} & \num{2.29e-01} & \num{2.60e-01} \\
    \hline\hline
    $\epsilon_\mathrm{expl}$ (erg) & \num{1.06e+51} [\num{1.09e+51}] & \num{1.06e+51} & \num{1.06e+51} & \num{1.06e+51} & \num{1.07e+51} & \num{1.09e+51} & \num{1.09e+51} \\
    \hline
    \end{tabular}
    \endgroup
    \tablefoot{
    \tablefoottext{a}{The values in the brackets refer to the results of a modified wd\_m090\_r0 simulation where we ignite the detionation with one tenth of the initial mass, i.e., $10^{-5}\,\msun$.}
    }
\end{table}

\begin{table}[h!]
    \centering
    \caption{Detailed abundance table for the models of the wd\_m100 series.}
    \label{tab:abundances_100}
    \begingroup
    \begin{tabular}{l|ccccccc}
    \hline
    {} &\multicolumn{7}{c}{wd\_m100}\\
    Yield ($\msun$) &  r0 &  r100 &  r200 &  r400 &  r800 &  r1600 &  r2800 \\
    \hline\hline
    $^{4}$He & \num{2.08e-03} & \num{2.09e-03} & \num{2.10e-03} & \num{2.13e-03} & \num{2.16e-03} & \num{2.45e-03} & \num{2.83e-03} \\
    $^{12}$C & \num{5.57e-03} & \num{5.58e-03} & \num{5.57e-03} & \num{5.56e-03} & \num{5.59e-03} & \num{5.75e-03} & \num{7.13e-03} \\
    $^{16}$O & \num{6.69e-02} & \num{6.69e-02} & \num{6.68e-02} & \num{6.67e-02} & \num{6.67e-02} & \num{6.75e-02} & \num{7.16e-02} \\
    $^{20}$Ne & \num{1.48e-03} & \num{1.49e-03} & \num{1.48e-03} & \num{1.48e-03} & \num{1.49e-03} & \num{1.63e-03} & \num{2.08e-03} \\
    $^{24}$Mg & \num{5.14e-03} & \num{5.13e-03} & \num{5.12e-03} & \num{5.12e-03} & \num{5.10e-03} & \num{5.27e-03} & \num{5.85e-03} \\
    $^{28}$Si & \num{2.16e-01} & \num{2.16e-01} & \num{2.16e-01} & \num{2.15e-01} & \num{2.14e-01} & \num{2.11e-01} & \num{2.03e-01} \\
    $^{32}$S & \num{1.33e-01} & \num{1.33e-01} & \num{1.33e-01} & \num{1.32e-01} & \num{1.32e-01} & \num{1.29e-01} & \num{1.24e-01} \\
    $^{36}$Ar & \num{2.57e-02} & \num{2.57e-02} & \num{2.56e-02} & \num{2.56e-02} & \num{2.54e-02} & \num{2.49e-02} & \num{2.39e-02} \\
    $^{40}$Ca & \num{2.43e-02} & \num{2.43e-02} & \num{2.42e-02} & \num{2.42e-02} & \num{2.41e-02} & \num{2.35e-02} & \num{2.26e-02} \\
    $^{44}$Ti & \num{1.92e-05} & \num{1.92e-05} & \num{1.91e-05} & \num{1.91e-05} & \num{1.92e-05} & \num{1.91e-05} & \num{1.89e-05} \\
    $^{48}$Cr & \num{4.99e-04} & \num{4.98e-04} & \num{4.96e-04} & \num{4.95e-04} & \num{4.94e-04} & \num{4.80e-04} & \num{4.65e-04} \\
    $^{52}$Fe & \num{1.09e-02} & \num{1.09e-02} & \num{1.08e-02} & \num{1.08e-02} & \num{1.08e-02} & \num{1.05e-02} & \num{1.01e-02} \\
    $^{56}$Ni & \num{4.84e-01} & \num{4.84e-01} & \num{4.84e-01} & \num{4.85e-01} & \num{4.87e-01} & \num{4.93e-01} & \num{5.00e-01} \\
    \hline\hline
    $\epsilon_\mathrm{expl}$ (erg)  & \num{1.35e+51} & \num{1.35e+51} & \num{1.35e+51} & \num{1.35e+51} & \num{1.35e+51} & \num{1.35e+51} & \num{1.34e+51} \\
    \hline
    \end{tabular}
    \endgroup
\end{table}

\begin{table}[h!]
    \centering
    \caption{Detailed abundance table for the models of the wd\_m110 series.}
    \label{tab:abundances_110}
    \begingroup
    \begin{tabular}{l|ccccccc}
    \hline
    {} &\multicolumn{7}{c}{wd\_m110}\\
    Yield ($\msun$) &  r0 &  r100 &  r200 &  r400 &  r800 &  r1600 &  r2800 \\
    \hline\hline
    $^{4}$He & \num{6.75e-03} & \num{6.75e-03} & \num{6.75e-03} & \num{6.74e-03} & \num{6.80e-03} & \num{6.89e-03} & \num{7.03e-03} \\
    $^{12}$C & \num{4.47e-03} & \num{4.48e-03} & \num{4.48e-03} & \num{4.48e-03} & \num{4.41e-03} & \num{4.51e-03} & \num{5.69e-03} \\
    $^{16}$O & \num{3.80e-02} & \num{3.80e-02} & \num{3.79e-02} & \num{3.76e-02} & \num{3.59e-02} & \num{3.66e-02} & \num{4.07e-02} \\
    $^{20}$Ne & \num{9.01e-04} & \num{9.01e-04} & \num{9.08e-04} & \num{8.91e-04} & \num{8.73e-04} & \num{9.20e-04} & \num{1.18e-03} \\
    $^{24}$Mg & \num{2.67e-03} & \num{2.67e-03} & \num{2.66e-03} & \num{2.63e-03} & \num{2.47e-03} & \num{2.56e-03} & \num{3.01e-03} \\
    $^{28}$Si & \num{1.44e-01} & \num{1.44e-01} & \num{1.44e-01} & \num{1.43e-01} & \num{1.36e-01} & \num{1.34e-01} & \num{1.32e-01} \\
    $^{32}$S & \num{9.17e-02} & \num{9.15e-02} & \num{9.15e-02} & \num{9.08e-02} & \num{8.69e-02} & \num{8.60e-02} & \num{8.42e-02} \\
    $^{36}$Ar & \num{1.84e-02} & \num{1.84e-02} & \num{1.84e-02} & \num{1.83e-02} & \num{1.76e-02} & \num{1.74e-02} & \num{1.70e-02} \\
    $^{40}$Ca & \num{1.78e-02} & \num{1.78e-02} & \num{1.78e-02} & \num{1.77e-02} & \num{1.72e-02} & \num{1.71e-02} & \num{1.66e-02} \\
    $^{44}$Ti & \num{2.09e-05} & \num{2.09e-05} & \num{2.09e-05} & \num{2.09e-05} & \num{2.07e-05} & \num{2.09e-05} & \num{2.09e-05} \\
    $^{48}$Cr & \num{4.08e-04} & \num{4.08e-04} & \num{4.08e-04} & \num{4.09e-04} & \num{4.08e-04} & \num{4.07e-04} & \num{3.98e-04} \\
    $^{52}$Fe & \num{8.60e-03} & \num{8.61e-03} & \num{8.62e-03} & \num{8.68e-03} & \num{8.79e-03} & \num{8.82e-03} & \num{8.62e-03} \\
    $^{56}$Ni & \num{7.19e-01} & \num{7.20e-01} & \num{7.20e-01} & \num{7.23e-01} & \num{7.36e-01} & \num{7.37e-01} & \num{7.36e-01} \\
    \hline\hline
    $\epsilon_\mathrm{expl}$ (erg) & \num{1.57e+51} & \num{1.57e+51} & \num{1.57e+51} & \num{1.57e+51} & \num{1.58e+51} & \num{1.58e+51} & \num{1.57e+51} \\
    \hline
    \end{tabular}
    \endgroup
\end{table}
\end{appendix}

\end{document}